\documentclass[times,12pt]{article}

\usepackage{times}
\usepackage{natbib}
\usepackage{graphicx}
\usepackage{amsmath}
\usepackage{amsthm}
\usepackage{amssymb}
\usepackage{setspace}
\usepackage[margin=1in]{geometry}
\RequirePackage[colorlinks,citecolor=blue,urlcolor=blue]{hyperref}

\usepackage{amsmath}
\usepackage{amsthm}
\usepackage{amssymb}
\RequirePackage{color,ae,fancyvrb}
\RequirePackage[T1]{fontenc}
\usepackage{caption}
\usepackage{subcaption}
\usepackage{lineno}
\usepackage{ulem}
\newcommand{\pkg}[1]{{\fontseries{b}\selectfont #1}}
\DefineVerbatimEnvironment{Sinput}{Verbatim}{fontshape=sl}
\let\proglang=\textsf

\newcommand{\bftheta}{{\boldsymbol \theta}}

\newcommand{\bfSig}{{\bf \Sigma}}
\newcommand{\bfs}{{\bf s}}
\newcommand{\bfx}{{\bf x}}

\begin{document}

\title{Nonstationary Bayesian modeling for a large data set of derived surface temperature return values}

\author{Mark D. Risser}
\date{}
\maketitle  

{\scriptsize
\hskip4ex Affiliation: Lawrence Berkeley National Laboratory

\hskip4ex Correspondence email: \tt{mdrisser@lbl.gov}
}

\vskip3ex

\begin{abstract}
Heat waves resulting from prolonged extreme temperatures pose a significant risk to human health globally. Given the limitations of observations of extreme temperature, climate models are often used to characterize extreme temperature globally, from which one can derive quantities like return values to summarize the magnitude of a low probability event for an arbitrary geographic location. However, while these derived quantities are useful on their own, it is also often important to apply a spatial statistical model to such data in order to, e.g., understand how the spatial dependence properties of the return values vary over space and emulate the climate model for generating additional spatial fields with corresponding statistical properties. For these objectives, when modeling global data it is critical to use a nonstationary covariance function. Furthermore, given that the output of modern global climate models can be on the order of $\mathcal{O}(10^4)$, it is important to utilize approximate Gaussian process methods to enable inference. In this paper, we demonstrate the application of methodology introduced in \cite{Risser2020} to conduct a nonstationary and fully Bayesian analysis of a large data set of 20-year return values derived from an ensemble of global climate model runs with over 50,000 spatial locations. This analysis uses the freely available \pkg{BayesNSGP} software package for \proglang{R}.
\end{abstract}

\begin{center}
\end{center}

\onehalfspacing

\section{Introduction} \label{section1}

While the impact of anthropogenic forcings like greenhouse gas emissions on the global climate system is not yet significant or detectable for many weather and climate phenomena, the human impact on surface air temperature is well established in the literature. In recent years this influence has been documented for extreme temperature events \citep{stott2016attribution}, which can cause significant health risks for humans \citep{wehner2016deadly}. Large ensembles of high-resolution global climate models \citep[e.g.][]{stone2016benchmark} are often used to study the underlying dynamical and theromdynamical aspects of extreme temperature and also quantify the anthropogenic influence on extreme temperature events, commonly termed probabilistic extreme event attribution \citep{national2016attribution}. These large ensembles can be used to estimate quantities associated with the climatological distribution of extreme temperature, e.g., return values or return periods \citep{coles2001}. However, while these derived quantities are useful on their own, it is also often important to apply a spatial statistical model to such data for several reasons: (1) to estimate return values for the grid cells where the extreme value analysis resulted in missing values, (2) to reduce the signal-to-noise ratio of the return value estimates by borrowing strength over space, (3) to understand how the spatial dependence properties of the return values vary over space, and (4) to emulate the climate model for generating additional spatial fields with corresponding statistical properties. 

Particularly for these latter two purposes, it is critical to utilize a nonstationary spatial statistical analysis \citep{Sampson1992,Stein2005,Paciorek2006}, where the spatial covariance properties of the underlying process of interest are allowed to vary over the globe. However, the output of global climate models is often very large, with a $1^\circ \times 1^\circ$ spatial grid resulting in over 50,000 grid cells. As such, it is important to utilize approximate Gaussian process methods \citep[e.g.][]{datta2016hierarchical,katzfuss2017multi,katzfuss2017general} in order to avoid the computational burdens associated with evaluating the exact Gaussian process likelihood. 

In this paper, we utilize methodology developed by \cite{Risser2020} to conduct a nonstationary and fully Bayesian analysis of a large data set of estimated return values from an ensemble of global climate model runs. In Section \ref{section2} we describe the climate model output used and the extreme value analysis used to estimate the return values. In Section \ref{section3} we outline a Bayesian nonstationary Gaussian process model for the return values, including a full description of the nonstationary covariance function, our approach for characterizing nonstationarity, and computational details for the Markov chain Monte Carlo algorithm required to fit the fully Bayesian model. Section \ref{section4} summarizes our findings, and reproducibility code is provided in Appendix \ref{appx}.

\section{Data} \label{section2}

\begin{figure}[!t]
\begin{center}
\includegraphics[trim={0 0 0 0mm}, clip, width = \textwidth]{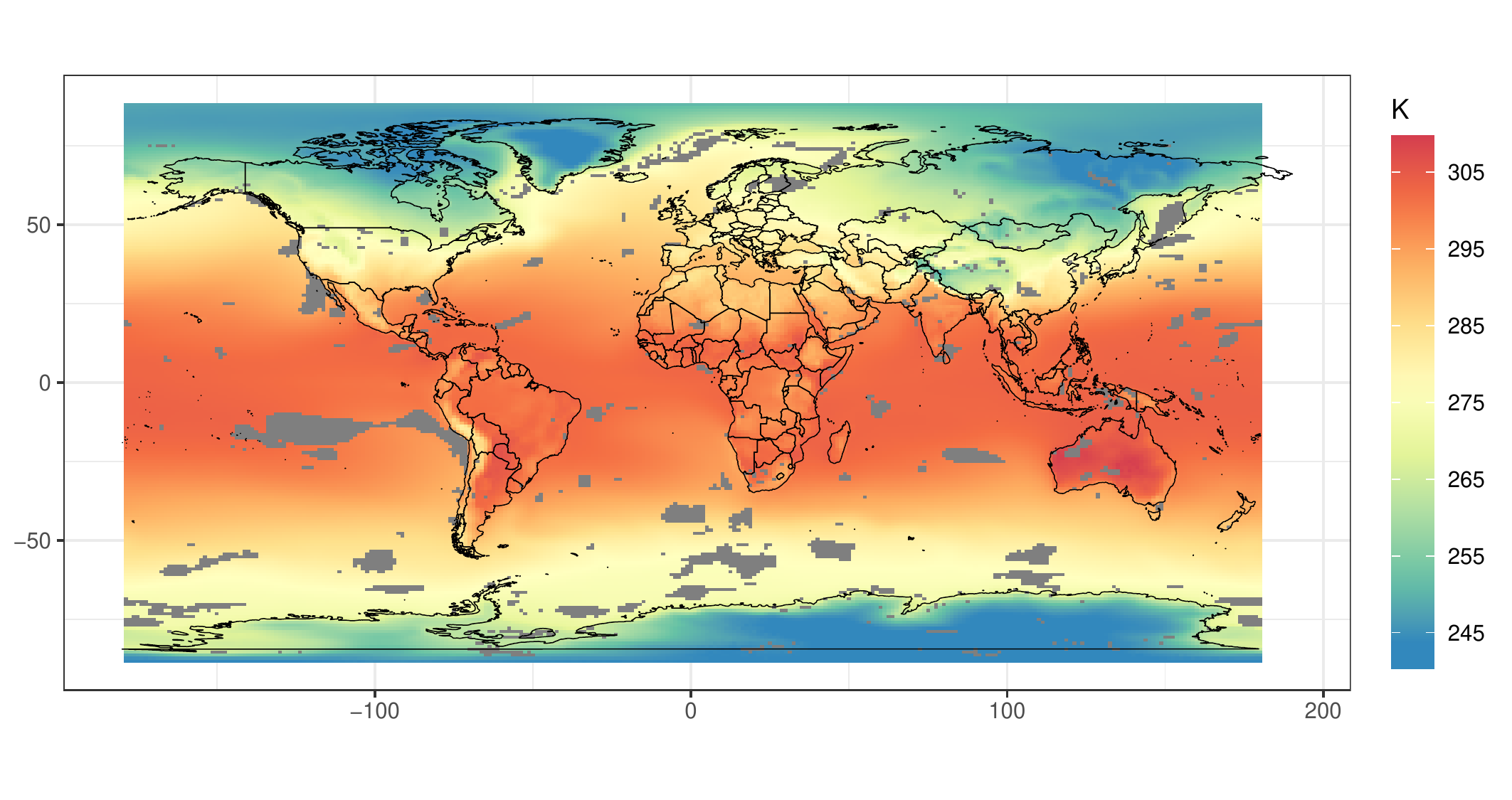}
\caption{Derived maximum likelihood estimates of the 20-year return value for DJF average temperature at the surface or TAS (in K), calculated independently for each grid cell using the ensemble maximum DJF average temperature in each year from the CAM5.1-1degree simulations over 1960-2015. Gray cells indicate grid cells where the maximum likelihood estimation failed.}
\label{figure3}
\end{center}
\end{figure}

The data used here are based on the output of an ensemble of climate model simulations from version 5.1 of the Community Atmospheric Model global atmosphere/land climate model, run in its conventional $\approx$1$^{\circ}$ longitude/latitude configuration \citep{NealeRB_ChenC-C_etalii_2012,stone2018basis}.  These simulations were run under the experiment protocols of the C20C+ Detection and Attribution Project \citep{stone2016benchmark} following two climate scenarios \citep{AngelilO_StoneD_etalii_2016}. We utilize simulations from the ``factual'' (or historical) scenario, which is driven by observed boundary conditions of atmospheric chemistry (greenhouse gases, tropospheric and stratospheric aerosols, ozone), solar luminosity, land use/cover, and the ocean surface (temperature and ice coverage).  The data and further details on the simulations are available at {\url{http://portal.nersc.gov/c20c}}; we use the 50-member ensemble that covers 01/1959 to 12/2014. A climate model ensemble is a set of runs from a particular climate model where each ensemble member has the same boundary conditions but stochastically perturbed initial conditions; the different ensemble members can be considered independent samples from the population defined by the climate model.   

This large ensemble of simulations is particularly suitable for evaluating climate extremes. As such, here we analyze derived maximum likelihood estimates of $r$-year return values for the DJF seasonal mean temperature over 1960-2015. The $r$-year return value corresponds to the seasonal mean temperature value that is expected to occur once every $r$ years, on average (for more information, we refer the interested reader to \citealp{coles2001}, chapter 3). These return values are calculated as follows: first, using the 50-member ensemble of simulations, for each grid cell we first extract the ensemble maximum DJF average temperature from each year, leaving us with a sample of 56 maxima; denote these maxima as $\{m_t(\bfs): \bfs \in G, t = 1960,\dots,2015\}$. For an individual grid cell $\bfs$, the cumulative distribution function (CDF) of the $\{m_t(\bfs): t = 1960,\dots,2015\}$ can be well-approximated by a member of the GEV family
\begin{equation} \label{gev_fam}
G_{\bfs}(x) \equiv \mathbb{P}(m_t(\bfs) \leq x) = \exp\left\{-\left[ 1 + \xi(\bfs)\left(\frac{x - \mu(\bfs)}{\sigma(\bfs)}\right) \right]^{-1/\xi(\bfs)} \right\}
\end{equation}
\cite[][Theorem~3.1.1, page~48]{coles2001}, defined for $\{ x: 1 + \xi(\bfs)(x - \mu(\bfs))/\sigma(\bfs) > 0 \}$. The GEV family of distributions~(\ref{gev_fam}) is characterized by three statistical parameters: the location parameter $\mu(\bfs) \in \mathcal{R}${, }which describes the center of the distribution{;} the scale parameter $\sigma(\bfs)>0$, which describes the spread of the distribution{;} and the shape parameter $\xi(\bfs) \in \mathcal{R}$. The shape parameter $\xi(\bfs)$ is the most important {for} determining the qualitative behavior of the distribution of daily rainfall at a given location{. I}f $\xi(\bfs)<0$, the distribution has a finite upper bound; if $\xi(\bfs) >0$, the distribution has no upper limit; {and} if $\xi(\bfs) = 0$, the distribution is again unbounded and the CDF~(\ref{gev_fam}) is interpreted as the limit $\xi(\bfs) \rightarrow 0$ \citep{coles2001}.

Next, we fit a Generalized Extreme Value distribution to these maxima using maximum likelihood estimation (using the \pkg{climextRemes} package; \citealp{R_climextRemes}), resulting in estimates $\big\{\widehat{\mu}(\bfs), \widehat{\sigma}(\bfs), \widehat{\xi}(\bfs) \big\}$ for all $\bfs$. These MLEs can be used to calculate corresponding estimates of the DJF $20$-year return value, denoted $z(\bfs)$, which is defined as the DJF average temperature that is expected to be exceeded on average once every 20 years. In other words, $z(\bfs)$ is an estimate of the $1-1/20$ quantile of the distribution of DJF maximum average temperature at grid cell $\bfs$, i.e.,
$\widehat{P}\big(m_{t}(\bfs) > z(\bfs)\big) = {1}/{20}$, which can be written in closed form in terms of the GEV parameters:
\begin{equation} \label{returnVal}
z(\bfs) = \left\{ \begin{array}{ll}
\widehat{\mu}(\bfs) - \frac{\widehat{\sigma}(\bfs)}{\widehat{\xi}(\bfs)}\big[1 - \{-\log(1-1/20)\}^{-\widehat{\xi}(\bfs)}\big],  & \widehat{\xi}(\bfs) \neq 0 \\[1ex]
\widehat{\mu}(\bfs) - \widehat{\sigma}(\bfs) \log\{-\log(1-1/20)\},  & \widehat{\xi}(\bfs) = 0.
\end{array} \right. 
\end{equation}
\citep{coles2001}. The default optimization procedure in \pkg{climextRemes} fails for some grid cells, in which case we record a missing value; otherwise, the estimated return values are given in Kelvin (K) and are considered fixed for the remainder of the analysis (see Figure \ref{figure3}; grid cells where the optimization failed are plotted in gray). The model is defined on a $288 \times 192$ global grid with $55,296$ grid cells, although we exclude the extreme pole model grid cells with a latitude of greater than $\pm 89^\circ$ leaving $54,144$ cells, $N=51,483$ of which are non-missing. 

\section{Nonstationary spatial modeling} \label{section3}

As described in Section \ref{section1}, a spatial model fit to these data is useful for several reasons: (1) to estimate return values for the grid cells with missing values, (2) to reduce the signal-to-noise ratio of the return value estimates by borrowing strength over space, (3) to understand how the spatial dependence properties of the return values vary over space, and (4) to emulate the climate model for generating additional spatial fields with corresponding statistical properties \citep[see, e.g.,][]{li2019efficient}. Particularly for the latter two purposes, for a global data set it is important to allow the properties of the covariance function to vary over space. 

We now outline a general statistical modeling framework for the derived return values (Section \ref{CGPM}) and then describe the nonstationary covariance model and prior distributions used (Section \ref{section32}). The approximate likelihood used to model this moderately large data set is then described (Section \ref{section33}; finally, we outline the Markov chain Monte Carlo used to fit the Bayesian model (Section \ref{section34}).

\subsection{Canonical Bayesian Gaussian process model} \label{CGPM}

In order to model the return values spatially, we use a univariate spatial Gaussian process: define $\{ z(\bfs) : \bfs \in G\}$ to be the MLE of the return values at grid cell $\bfs$, where $G \subset \mathcal{R}^3$ is the set of three dimensional coordinates of the globe. The statistical model can be written as 
\begin{equation} \label{CANONmodel}
z({\bf s}) =  y({\bf s}) + \varepsilon({\bf s}), 
\end{equation}
where $E[z({\bf s})] = y(\bfs)$, $y(\cdot)$ is a spatial random effect, and $\varepsilon(\cdot)$ is a stochastic component that represents measurement error or microscale variability and is independently distributed as ${N}(0, \tau^2(\bfs))$ such that $\varepsilon(\cdot)$ and $y(\cdot)$ are independent. The spatial random effect is modeled as a parametric Gaussian process, denoted $y(\cdot) \sim GP\big( \mu, C_y(\cdot,\cdot; \bftheta_y)\big)$, such that $E[y(\bfs)] = \mu$. The covariance function $C_y$ is assumed known up to a vector of parameters $\bftheta_y$ and describes the covariance between the process $y(\cdot)$ as  
\[
C_y(\bfs, \bfs'; \bftheta_y) \equiv Cov\big(y(\bfs), y(\bfs') \big),
\]
for all $\bfs, \bfs' \in G$. Finally, we suppose that the error variance process $\tau^2(\cdot)$ is known up to a vector of parameters $\bftheta_z$.

Define $\mathcal{S}_O = \{{\bf s}_1, ... , {\bf s}_N\}\in G$ (recall $N=51,483$) to be the spatial locations where we have a non-missing return value (i.e., grid cells where the GEV optimization did not fail); (\ref{CANONmodel}) implies that the observed vector ${\bf z}_O = \left[ z({\bf s}_1), ... , z({\bf s}_N) \right]^\top$ has a multivariate Gaussian distribution 
\begin{equation} \label{Zcond}
p({\bf z}_O | {\bf y}_O, \bftheta_z ) = {N}\big( {\bf y}_O, {\bf \Delta}(\bftheta_z)\big),
\end{equation}
where ${\bf \Delta}(\bftheta_z) = diag[\tau^2(\bfs_1), \dots, \tau^2(\bfs_N)]$. Conditional on the other parameters in the model, the process vector ${\bf y}_O = \left[ y({\bf s}_1), ... , y({\bf s}_N) \right]^\top$ is distributed as 
\begin{equation} \label{Ycond}
p({\bf y}_O | \mu, \bftheta_y) = {N}\big( \mu {\bf 1}_N, {\bf \Omega}(\bftheta_y) \big),
\end{equation}
where the elements of ${\bf \Omega}(\bftheta_y)$ are $\Omega_{ij} \equiv C_y({\bf s}_i, {\bf s}_j; \bftheta_y)$. Given the Gaussian distributions in (\ref{Zcond}) and (\ref{Ycond}), it is often useful to integrate over the process $y(\cdot)$ to arrive at the marginal distribution for $z(\cdot)$, which is
\begin{equation} \label{Zmarg}
p({\bf z}_O | \mu, \bftheta) = \int p({\bf z}_O |  {\bf y}_O, \bftheta_z ) p({\bf y}_O | \mu,\bftheta_y) d{\bf y}_O = {N}\big(\mu {\bf 1}_N, {\bf \Delta}(\bftheta_z) + {\bf \Omega}(\bftheta_y) \big)
\end{equation}
where $\bftheta = (\bftheta_z, \bftheta_y)$. The covariance function for the marginalized process is
\begin{equation} \label{Zcov}
C_z(\bfs, \bfs'; \bftheta) = C_y(\bfs, \bfs'; \bftheta_y) + \tau(\bfs)\tau(\bfs')I_{\{\bfs = \bfs'\}}, \hskip4ex \text{for all } \bfs, \bfs' \in G,
\end{equation}
where $I_{\{\cdot\}}$ is an indicator function. 

To complete the Bayesian specification of this model, we define prior distributions for the unknown mean and covariance parameters $p(\mu, \bftheta)$, where these priors are assumed to be independent (i.e., $p(\mu, \bftheta) = p(\mu) p(\bftheta)$) and noninformative: we assume $p(\mu) = N(0, 100^2)$; see Section \ref{section32} for more details on the priors used for $p(\bftheta)$. All inference for $\mu$ and $\bftheta$ is based on the marginalized posterior for these parameters conditional on ${\bf z}_O$:
\begin{equation} \label{posteriorZ}
p(\mu, \bftheta | {\bf z}_O) \propto p({\bf z}_O | \mu, \bftheta) p(\mu) p(\bftheta).
\end{equation}
Regardless of the form of the priors on $\mu$ and $\bftheta$, the posterior distribution (\ref{posteriorZ}) is not available in closed form, and so we must resort to Markov chain Monte Carlo (MCMC) methods to conduct inference on $\mu$ and $\bftheta$. See Section \ref{section34} for more information on the MCMC.

Posterior prediction of the true unobserved return values $y(\bfs)$ for both $\mathcal{S}_O$ (the locations with non-missing return values) and $\mathcal{S}_P = \{{\bf s}^*_1, ... , {\bf s}^*_M\}$ (the locations with missing return values; here $M=2,661$) is straightforward given the Gaussian process assumptions used here. Define ${\bf y}_P= \left( y({\bf s}^*_1), ... , y({\bf s}^*_M) \right)^\top$ and ${\bf y} = ({\bf y}_O, {\bf y}_P)$; the predictive distribution of interest is then
\begin{equation} \label{ppd}
p({\bf y} | {\bf z}_O ) = \int_{\mu, \bftheta} p({\bf y}, \mu, \bftheta | {\bf z}_O) d\mu d\bftheta = \int_{\mu, \bftheta} p({\bf y}| \mu, \bftheta, {\bf z}_O) p(\mu, \bftheta | {\bf z}_O) d\mu d\bftheta.
\end{equation}
The first component inside the integral on the far right hand side of (\ref{ppd}), i.e., $p({\bf y}| \mu, \bftheta, {\bf z}_O)$, is available in closed form. The Gaussian process assumption yields that the joint distribution of $({\bf z}_O, {\bf y})$ conditional on $(\mu, \bftheta)$ is
\begin{equation*}
p\left( \left. \left[ \begin{array}{c} {\bf z}_O \\ {\bf y} \end{array} \right] \right| \mu, \bftheta \right) = N\left(  \left[ \begin{array}{c} \mu {\bf 1}_N \\ \mu {\bf 1}_{N+M} \end{array} \right], \left[ \begin{array}{cc} {\bf C}_{{\bf z}_O} & {\bf C}_{{\bf z}_O, {\bf y}}\\ {\bf C}_{{\bf y}, {\bf z}_O}& {\bf C}_{\bf y}\end{array} \right] \right)
\end{equation*}
(implicit conditioning on $\bftheta$ on the right hand side suppressed), where the entries of the covariance (or cross-covariance) matrices ${\bf C}_{(\cdot)}$ are determined from the covariance functions $C_{(\cdot)}$. Based on the conditional properties of the multivariate Gaussian distribution, we have
\begin{equation*}
p\left(  {\bf y} | {\bf z}_O, \mu, \bftheta \right) = N\left( {\bf m}_{{\bf y}|{\bf z}_O}, {\bf C}_{{\bf y}|{\bf z}_O} \right),
\end{equation*}
where
\[
{\bf m}_{{\bf y}|{\bf z}_O} = \mu {\bf 1}_{N+M} + {\bf C}_{{\bf y}, {\bf z}_O} {\bf C}_{{\bf z}_O}^{-1} ({\bf z}_O - \mu {\bf 1}_{N})
\]
and
\[
{\bf C}_{{\bf y}|{\bf z}_O} = {\bf C}_{\bf y} - {\bf C}_{{\bf y}, {\bf z}_O} {\bf C}_{{\bf z}_O}^{-1} {\bf C}_{{\bf z}_O, {\bf y}}.
\]
The other component under the integral on the far right hand side of (\ref{ppd}) is the posterior (\ref{posteriorZ}); hence, in practice, given a set of posterior samples $\{ \mu_l, \bftheta_l : l = 1, \dots, L \}$ (obtained via MCMC), a Monte Carlo estimate of (\ref{ppd}) is obtained via
\[
p({\bf y} | {\bf z}_O ) \approx \sum_{l=1}^L p\left(  {\bf y} | {\bf z}_O, \mu_l, \bftheta_l \right).
\]

\subsection{Specifying nonstationarity} \label{section32}

In order to characterize heterogeneity in the second-order properties of the return values, it is important to use a nonstationary covariance function for $C_y$. Among the diverse literature on approaches for modeling a nonstationary covariance function, one of the more intuitive methods involves allowing the parameters of the covariance function $C_y$ to vary over space, the so-called spatially-varying parameters approach. Following \cite{Paciorek2006}, \cite{Risser2015}, and many others, one approach for modeling the parametric covariance function for $y(\cdot)$ is via 
\begin{equation} \label{PScov}
C_y(\bfs, \bfs'; \bftheta) = \sigma(\bfs) \sigma(\bfs') \frac{\left|\bfSig(\bfs)\right|^{1/4}\left|\bfSig(\bfs')\right|^{1/4}}{\left|\frac{\bfSig(\bfs') + \bfSig(\bfs')}{2} \right|^{1/2}} \mathcal{M}_\nu\left(\sqrt{Q(\bfs, \bfs')}\right), \hskip3ex \bfs, \bfs' \in G
\end{equation}
where
\begin{equation} \label{Qij}
Q(\bfs, \bfs') = (\bfs - \bfs')^\top \left(\frac{\bfSig(\bfs) + \bfSig(\bfs')}{2}\right)^{-1}(\bfs - \bfs')
\end{equation}
and $\mathcal{M}_\nu(\cdot)$ is the Mat\'ern correlation function with smoothness $\nu$ (note, however, that $C_y$ is non-negative definite for any valid correlation function over $\mathcal{R}^d$, $d \geq 1$). In (\ref{PScov}), $\sigma(\cdot)$ is a spatially-varying standard deviation process and $\bfSig(\cdot)$ is a spatially-varying anisotropy process that controls the range and direction of dependence. The covariance function in Eq.~\ref{PScov} arises as a generalization of a kernel convolution-based approach for constructing a positive definite covariance function \cite[e.g.,][]{Higdon1999}; see \cite{Paciorek2006} for further details. The covariance function defined via (\ref{Zcov}) and (\ref{PScov}) is highly flexible, as it defines parameter processes $\sigma(\cdot)$ and $\bfSig(\cdot)$---and $\tau(\cdot)$, the standard deviation process for the error $\varepsilon(\cdot)$, when considering $C_z$---over an infinite-dimensional space (i.e., $G \subset \mathcal{R}^d$). In practice, these processes must be regularized somehow so that implementation is feasible. To this end, we utilize the framework described in \cite{Risser2020}, where parametric regression models are used to regularize the spatial parameter fields.

Looking at the data in Figure \ref{figure3}, it is clear that variability of the return values exhibit strong zonal behavior (i.e., similar behavior for a fixed latitude), which is common for a global atmospheric variable like seasonal temperature. For the spatial variance, as a simple way of characterizing spatial variability in the covariance function we specify that the log of $\sigma(\cdot)$ varies nonparametrically with latitude via a set of natural splines with three degrees of freedom. However, looking at the data, it appears as though the return values have different properties over the land versus the ocean; hence, we assign separate coefficients to the natural splines for grid cells over the ocean and grid cells over land. This statistical model can be written as 
\begin{equation} \label{LLR}
\log \sigma(\bfs) = \bfx_\sigma(\bfs)^\top \boldsymbol{\alpha}, 
\end{equation}
where $\boldsymbol{\alpha}$ is a vector of regression coefficients and $\bfx_\sigma(\bfs)$ is a row of the design matrix specified using natural splines (with \texttt{ns(latitude, df = 3)} in \pkg{R}) interacted with a variable that indicates whether the grid cell is over land; this design matrix has an intercept with three columns for the natural spline functions, plus an additional four columns for the interaction. (Note: in the \cite{Risser2020} framework, this model can be specified using \texttt{"logLinReg"}.) Next, we specify that the ansiotropy process is different for ocean versus land, but a spatial constant for all ocean grid cells and a different spatial constant for all land grid cells. Since the covariance function (\ref{PScov}) is not valid on the sphere, we instead represent the longitude/latitude coordinates as points in a three-dimensional space and therefore must use the locally isotropic version of (\ref{PScov}). In other words, we force the covariance to be locally isotropic by setting the anisotropy process to be equal to a multiple of the identity matrix:
\[
\bfSig(\bfs) \equiv \Sigma(\bfs){\bf I}_3,
\]
where $\Sigma(\bfs)$ is a scalar. Then, similar to $\sigma(\cdot)$, we use a log-linear regression framework wherein
\begin{equation} \label{LLR2}
\log \Sigma(\bfs) = \bfx_\Sigma(\bfs)^\top \boldsymbol{\phi}, 
\end{equation}
where $\boldsymbol{\phi}$ is a vector of regression coefficients and $\bfx_\Sigma(\bfs)$ is a row of the design matrix constructed using a covariate that indicates whether the grid cell is over land or over the ocean. (Note: in the \cite{Risser2020} framework, this model can be specified using \texttt{"compRegIso"}.) Otherwise, given that we are dealing with gridded data that are relatively smooth, we specify the nugget variance to be an unknown spatial constant (i.e., $\tau^2(\bfs) \equiv \tau^2$). Furthermore, we maintain a constant spatial mean to emphasize the second-order properties of the return values (i.e., $\mu(\bfs) \equiv \mu$). Since $\boldsymbol{\alpha}\in\mathcal{R}^8$ and $\boldsymbol{\phi}\in\mathcal{R}^2$, this results in a total of 12 covariance parameters.

Using this statistical model, $\bftheta = (\mu, \tau^2, \boldsymbol{\alpha}, \boldsymbol{\phi})$. The prior distributions used are independent, i.e.,
\[
p(\bftheta) = p(\mu) \times p(\tau^2) \times p(\boldsymbol{\alpha}) \times p(\boldsymbol{\phi}),
\]
as well as non-informative and diffuse:
\begin{equation}
\begin{array}{c}
     p(\mu) = N(0, 100^2)  \\
     p(\tau^2) = U(0, 100) \\
     p(\boldsymbol{\alpha}) = N_8({\bf 0}, 10^2{\bf I}) \\
     p(\boldsymbol{\phi}) = N_2({\bf 0}, 5^2{\bf I}) \cdot I_{\{ \max{\Sigma(\bfs)} < 12,742 \}}.
\end{array}
\end{equation}
The indicator function in $p(\boldsymbol{\phi})$ is for identifiability purposes; this ensures that the local isotropic range values do not exceed 12,742km (the diameter of Earth).

\subsection{Approximate GP inference and prediction} \label{section33}

Despite the fact that Gaussian processes are mathematically convenient representations for a spatial process and that prediction is straightforward, numerical calculations regarding the multivariate Gaussian distribution needed in evaluating (\ref{Zmarg}) for $N$ spatial locations require $\mathcal{O}(N^2)$ memory and $\mathcal{O}(N^3)$ time complexity. This is an issue for any application of Gaussian processes, but is particularly problematic for modeling nonstationary covariance functions which involve high-dimensional parameter spaces. However, when dealing with large data sets, we can utilize the diverse literature on approximate Gaussian process methods \citep[see][for a review and comparison of existing approaches]{heaton2018case} to make parameter inference and prediction feasible. 

The nearest-neighbor Gaussian process (NNGP; \citealp{datta2016hierarchical}) is one specific method that enables large data inference via Gaussian processes by forcing the precision matrix to be sparse. Actually, both of these methods can be framed as special cases of Vecchia approximations of Gaussian processes. \cite{Risser2020} outline a specific framework for implementing the nearest neighbor Gaussian process for the response \citep[NNGP-R,][]{finley2018efficient}; here, the NNGP-R approximation is applied to yield a sparse Cholesky factor of $\text{Cov}({\bf z}_O)$, which is based on $C_z$ (from Eq.~\ref{Zcov}). \cite{finley2018efficient} derive a closed form expression for calculating the Cholesky factor of $\text{Cov}({\bf z}_O)$ and the subsequent quadratic forms needed to evaluate the likelihood. When the number of nonzero elements in the Cholesky decomposition is limited to $k$ (using a $k$ nearest neighbor scheme based on maxmin ordering), the Cholesky is guaranteed to be sparse and can be calculated by solving $N-1$ linear systems of size at most $k\times k$, which can be performed in $\mathcal{O}(Nk^3)$ flops (furthermore, parallelization could be employed, since each linear system can be solved independently of all others). As such, the likelihood can then be calculated in $\mathcal{O}(Nk)$ time complexity \citep{finley2018efficient}, which is linear in $N$. Using NNGP-R,  posterior prediction can only be accomplished for individual locations (also called ``local kriging'') because the covariance corresponding to the prediction locations is diagonal \citep[Section 5.2.1 of][]{katzfuss2018vecchia}. \cite{finley2018efficient} outline an algorithm for posterior prediction of the response (i.e., $z(\cdot)$) at a single location \citep[Algorithm 4,][]{finley2018efficient}; \cite{katzfuss2018vecchia} note that the same framework can be used to predict either $z(\cdot)$ or $y(\cdot)$ by including or not including the nugget variance, respectively, in the prediction variance. 

\subsection{Markov chain Monte Carlo} \label{section34}

While this model does not contain too many parameters (12 total), it is still important to use block sampling to account for the correlated parameter space. Functionality provided by the \pkg{BayesNSGP} \citep{BayesNSGP} and \pkg{nimble} \citep{nimble_jcgs} packages for \proglang{R} makes customizing samplers for a general MCMC straightforward. For this application, we apply a single block Metropolis Hastings sampler for the eight spatial variance coefficients $\boldsymbol{\alpha}$ and another block Metropolis Hastings sampler to the two isotropy range coefficients $\boldsymbol{\phi}$. We maintain univariate adaptive random walk Metropolis Hastings samplers for the constant spatial mean and nugget variance. Looking at trace plots, the MCMC converges relatively quickly: we run the MCMC for 20,000 iterations, discarding the first 10,000 samples as burn-in and saving every 5th sample for posterior summaries and prediction. Computational times for each component of the model fitting and predictions are 37.3 hours and 13.4 hours, respectively (when running the analysis on one core of a 12-core Intel Xeon CPU E5520 machine with 128 GB memory). Note that the amount of time it takes the build and compile the model is somewhat non-trivial for this very large data set (on the order of 20 minutes). Furthermore, even with the NNGP-R likelihood, the MCMC is still somewhat slow.

The code used to generate this statistical model and conduct the MCMC and prediction is provided in Appendix \ref{appx}.

\section{Results} \label{section4}

\begin{figure}[!t]
\begin{center}
\includegraphics[trim={0 0 0 0mm}, clip, width = 0.85\textwidth]{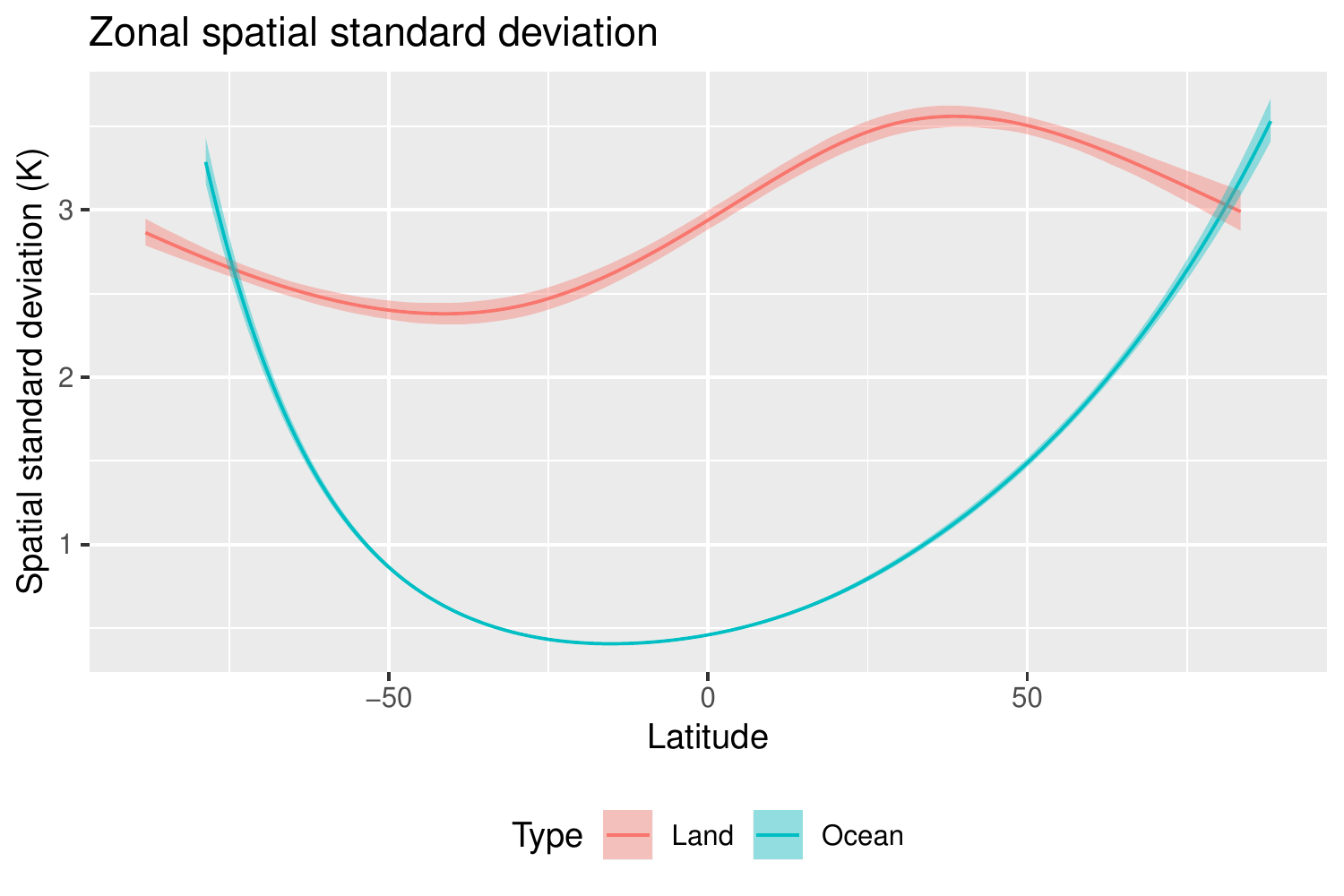}
\caption{Posterior means and 99\% Bayesian credible intervals (BCIs) for the zonal spatial standard deviations for grid cells over land versus ocean.}
\label{app3_post}
\end{center}
\end{figure}
 
The posterior mean for the overall mean is $297.94$ K, with a 99\% Bayesian credible interval (BCI) of $(297.70, 298.20)$; similarly, for the overall nugget standard deviation, the posterior mean is $7.2\times 10^{-7}$ K with a 99\% BCI of $(9.86\times 10^{-9}, 3.14 \times 10^{-6})$. The posterior means for the ocean/land isotropy range are 12738 km and 1321 km, respectively (with 99\% BCIs $(12726, 12741)$ and $(1306, 1334)$); clearly, the return values are much smoother over the ocean relative to the land. Recall, however, that the prior on the isotropy range has an upper bound of 12742 km (the diameter of the globe), such that the range over the ocean is very close to its upper bound. Finally, the zonal spatial standard deviation is shown in Figure \ref{app3_post}. In general, ocean regions have a much smaller spatial variance relative to the land (except at the poles), with particularly small variability in the tropics (within $\pm25^\circ$ latitude), and there are significant differences between ocean and land.

\begin{figure}[!t]
\begin{center}
\includegraphics[trim={0 0 0 0mm}, clip, width = \textwidth]{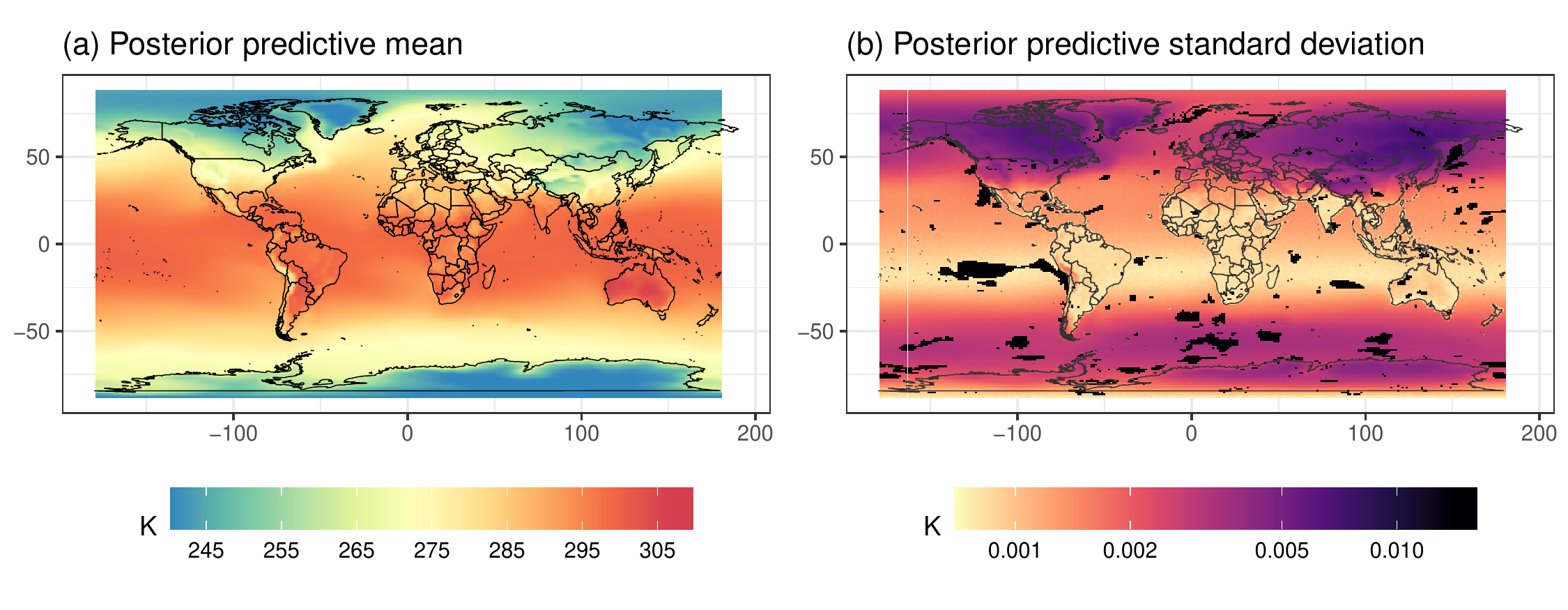}
\caption{Posterior predictive mean and standard deviation for the 20-year return values shown in Figure \ref{figure3}.}
\label{app3_pred}
\end{center}
\end{figure}

Next, we show the posterior predictive mean and standard deviation (both in K) in Figure \ref{app3_pred}. Unsurprisingly, given the very small estimated nugget standard deviation ($\approx 7.2\times 10^{-7}$K), the predictive mean looks very similar to the raw data given in Figure \ref{figure3}, with one important difference: the missing values have now been filled in via the local kriging enabled by the NNGP likelihood. Interestingly, the posterior standard deviation is very clearly a function of the proximity to non-missing data values (i.e., the grid cells with a missing data value have much larger uncertainty than those with a non-missing value) but also ocean versus land. 

\section*{Acknowledgements}

This work was supported by the Regional and Global Model Analysis Program of the Office of Biological and Environmental Research in the Department of Energy Office of Science under contract number DE-AC02-05CH11231. This document was prepared as an account of work sponsored by the United States Government. While this document is believed to contain correct information, neither the United States Government nor any agency thereof, nor the Regents of the University of California, nor any of their employees, makes any warranty, express or implied, or assumes any legal responsibility for the accuracy, completeness, or usefulness of any information, apparatus, product, or process disclosed, or represents that its use would not infringe privately owned rights. Reference herein to any specific commercial product, process, or service by its trade name, trademark, manufacturer, or otherwise, does not necessarily constitute or imply its endorsement, recommendation, or favoring by the United States Government or any agency thereof, or the Regents of the University of California. The views and opinions of authors expressed herein do not necessarily state or reflect those of the United States Government or any agency thereof or the Regents of the University of California.

\bibliographystyle{apalike} 
\bibliography{BayesNSGP}

\appendix
\section{Code to reproduce the analysis} \label{appx}

What follows is \proglang{R} code to build and compile the MCMC and conduct posterior prediction using the \pkg{BayesNSGP} package.

\begin{verbatim}
# Load required packages
library(nimble)
library(splines)
library(BayesNSGP)

#================================================
# Load data and setup
#================================================
tasRV20_DF_all <- read.csv("C20C_DJFtasRV20_trend.csv")
tasRV20_DF_all$ind_land <- as.factor(tasRV20_DF_all$ind_land)
# Trim extreme poles
tasRV20_DF_all <- tasRV20_DF_all[abs(tasRV20_DF_all$latitude) < 89,]

# Design matrices
Xmat_sigma <- model.matrix(~ ns(latitude, df = 3)*ind_land, data = tasRV20_DF_all)
Xmat_Sigma <- model.matrix(~ ind_land, data = tasRV20_DF_all)

# Remove NA's for fitting
tasRV20_DF <- tasRV20_DF_all[!is.na(tasRV20_DF_all$rv20),]
Xmat_sigma_train <- model.matrix(~ ns(latitude, df = 3)*ind_land, data = tasRV20_DF)
Xmat_Sigma_train <- model.matrix(~ ind_land, data = tasRV20_DF)

# Convert lon/lat to x/y/z
xyz.crds <- matrix(NA,nrow(tasRV20_DF),3)
# Transform degrees to radians
lat.radians <- tasRV20_DF$latitude*(pi/180)
lon.radians <- tasRV20_DF$longitude*(pi/180)
for(i in 1:nrow(xyz.crds)){
  # Earth radius ~ 6371km
  xyz.crds[i,1] <- 6.371*cos(lat.radians[i])*cos(lon.radians[i])
  xyz.crds[i,2] <- 6.371*cos(lat.radians[i])*sin(lon.radians[i])
  xyz.crds[i,3] <- 6.371*sin(lat.radians[i])
}

# Constants for NNGP 
constants <- list( 
  nu = 0.5, k = 15, mu_HP1 = 100, X_sigma = Xmat_sigma_train, sigma_HP1 = 10,
  X_Sigma = Xmat_Sigma_train, Sigma_HP1 = 5, maxAnisoRange = 2*6.371
)

#================================================
# MCMC using the NNGP likelihood
#================================================
# Set up the model
Rmodel <- nsgpModel(likelihood = "NNGP", constants = constants, 
    coords = round(xyz.crds, 4), data = tasRV20_DF$rv20, 
    tau_model = "constant", sigma_model = "logLinReg", 
    mu_model = "constant", Sigma_model = "compRegIso")
# Configure the MCMC
conf <- configureMCMC(Rmodel)
conf$removeSamplers(c("alpha[1:8]","Sigma_coef1[1:2]"))
conf$addSampler(target = c("alpha[1:8]"), type = "RW_block", silent = TRUE )
conf$addSampler(target = c("Sigma_coef1[1:2]"), type = "RW_block", silent = TRUE )

Rmcmc <- buildMCMC(conf) # Build MCMC
Cmodel <- compileNimble(Rmodel) # Compile the model
Cmcmc <- compileNimble(Rmcmc, project = Rmodel) # Compile the MCMC

# Initial values
initsList <- list( Sigma_coef1 = rep(0,2),
  alpha = rep(0,8), beta = 285, delta = 0.005
)

# Run
samples <- runMCMC(Cmcmc, niter = 20000, nburnin = 0, inits = initsList)

#================================================
# Prediction
#================================================
# Convert lon/lat to x/y/z
xyz.crds <- matrix(NA,nrow(tasRV20_DF_all),3)
# Transform degrees to radians
lat.radians <- tasRV20_DF_all$latitude*(pi/180)
lon.radians <- tasRV20_DF_all$longitude*(pi/180)
for(i in 1:nrow(xyz.crds)){
  xyz.crds[i,1] <- 6.371*cos(lat.radians[i])*cos(lon.radians[i]) 
  xyz.crds[i,2] <- 6.371*cos(lat.radians[i])*sin(lon.radians[i])
  xyz.crds[i,3] <- 6.371*sin(lat.radians[i])
}
pred_NNGP <- nsgpPredict(model = Rmodel, 
    samples = samples[seq(from=10005,to=20000,by=5),],
    coords.predict = round(xyz.crds, 4), PX_sigma = Xmat_sigma, 
    PX_Sigma = Xmat_Sigma)
\end{verbatim}

\end{document}